\documentclass[usenatbib, usegraphicx]{mn2e}
\bibpunct{(}{)}{;}{a}{}{,}
\def\cm2{cm$^{-2}$}
\def\hi{H\,{\sc i}}
\def\hii{H\,{\sc ii}}
\def\di{D\,{\sc i}}
\def\k3{Si\,{\sc iii}}
\def\s2{Si\,{\sc ii}}
\def\si4{Si\,{\sc iv}}
\def\fe2{Fe\,{\sc ii}}
\def\mg2{Mg\,{\sc ii}}
\def\c4{C\,{\sc iv}}
\def\carb2{C\,{\sc ii}}
\def\o1{O\,{\sc i}}

\newcommand{\lya}{Ly$\alpha$}
\newcommand{\lyb}{Ly$\beta$}
\newcommand{\kms}{~kms$^{-1}$}

\title{D/H in a new Lyman limit absorption system at $z=3.256$ towards
PKS1937-1009}

\author[] {\parbox[t]{\textwidth}
	{N. H. M. Crighton$^1$\thanks{E-mail: nhmc@phys.unsw.edu.au}, 
	J. K. Webb$^1$, 
	A. Ortiz-Gil$^2$, 
	A. Fern\'{a}ndez-Soto$^2$}
\\
\vspace*{6pt}\\
$^1$ School of Physics, University of New South Wales, Sydney NSW 2052, 
Australia \\
$^2$ Observatori Astron\`{o}mic, Universitat de Val\`{e}ncia, Burjassot
(Val\`{e}ncia), E-46100, Spain} 
\date{}

\pagerange{\pageref{firstpage}--\pageref{lastpage}}
\pubyear{2004}

\begin{document}

\maketitle

\label{firstpage}

\begin{abstract}

We have identified a new Lyman limit absorption system towards
PKS1937-1009, with log$N$(\hi) $=18.25 \pm 0.02$ at $z=3.256$ that is
suitable for measuring D/H.  We find a $68.3$\% confidence range for
D/H of $1.6^{+0.25}_{-0.30} \times 10^{-5}$, and a $95.4$\% range of
$1.6^{+0.5}_{-0.4} \times 10^{-5}$. The metallicity of the cloud where
D/H was measured is low, [Si/H] $= -2.0 \pm 0.5$. At these
metallicities we expect that D/H will be close to the primordial
value. Our D/H is lower than the D/H value predicted using the
$\Omega_b$ calculated from the cosmic background radiation measured by
WMAP, $2.60^{+0.19}_{-0.17} \times 10^{-5}$.  Our result also
exacerbates the scatter in D/H values around the mean primordial D/H.

\end{abstract}

\begin{keywords}
nuclear reactions, nucleosynthesis, abundances - quasars: absorption
lines - cosmological parameters.
\end{keywords}

\section{Introduction}

There are interesting disagreements between the various ways of
calculating the baryon density of the universe ($\Omega_b$).  The most
precise $\Omega_b$ measurement is from the recent WMAP observations of
the CMB power spectrum \citep{Spergel03}.  Another way to measure
$\Omega_b$ is by measuring the relative abundances of the light
elements produced in big bang nucleosyntheis (BBN), since in standard
BBN theory $\Omega_b$ is the only free parameter determining these
abundances.  Attempts have been made to measure the primordial $^4$He,
$^7$Li and deuterium (D) abundances.  In principle, $\Omega_b$ is most
reliably determined by D, due to its apparently uncomplicated
evolution with time \citep*{Epstein74} and because it depends more
sensitively on $\Omega_b$ than $^4$He and $^7$Li. There is broad
agreement between the D $\Omega_b$ and CMB $\Omega_b$, but the $^4$He
and $^7$Li measurements predict a lower $\Omega_b$ \citep{Coc04,
Cyburt03}.  Either there is something wrong with our interpretation of
the $^4$He and $^7$Li measurements, or the standard BBN theory may
need to be modified \citep[e.g. see][]{Barger03, Dmitriev04,
Ichikawa04}.

The mean of the primordial D values as measured in quasar absorption
systems is consistent with the WMAP CMB $\Omega_b$.  However, there is
statistically significant scatter of the individual D/H measurements
about the mean.  The metallicity of the absorption clouds used to make
the D/H measurements is low, [M/H]$< -1.5$, so that no astration (D
destruction due to stellar nucleosynthesis) is thought to have
occurred \citep{Fields96}.  In addition, the primordial D/H is thought
to be isotropic and homogeneous, so the scatter is hard to explain.
One explanation is that the systematic errors in measuring D/H have
been underestimated by the authors.  However, with only half a dozen
quasar absorption system D/H measurements other explanations, such as
some early mechanism for astration or a non standard BBN, cannot be
ruled out.  More D/H measurements are needed to resolve this problem.
It is exciting to see further results being added to this short list:
in the last year two D/H measurements have been published: one towards
another QSO absorption line system \citep{Kirkman03} and another in a
low metallicity gas cloud near our Galaxy \citep{Sembach03}.

\begin{figure}
\begin{center}
\includegraphics[width=0.49\textwidth]{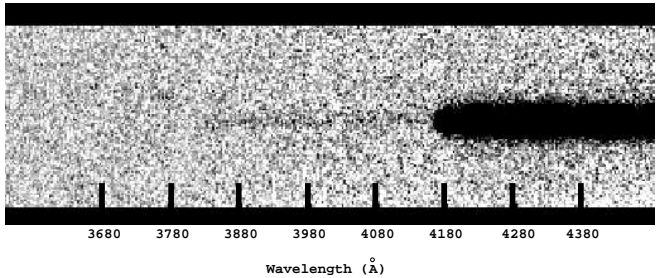}
\caption{\label{img} The two dimensional LRES spectrum, showing the
position of the $z=3.572$ LL at $\sim 4170$ \AA.  The LL due to the $z
= 3.256$ system we are interested in can be seen at $\sim 3880$ \AA.}
\end{center}
\end{figure}

\begin{figure*}
\begin{minipage}{140mm}
\begin{center}
\includegraphics[width=1.0\textwidth]{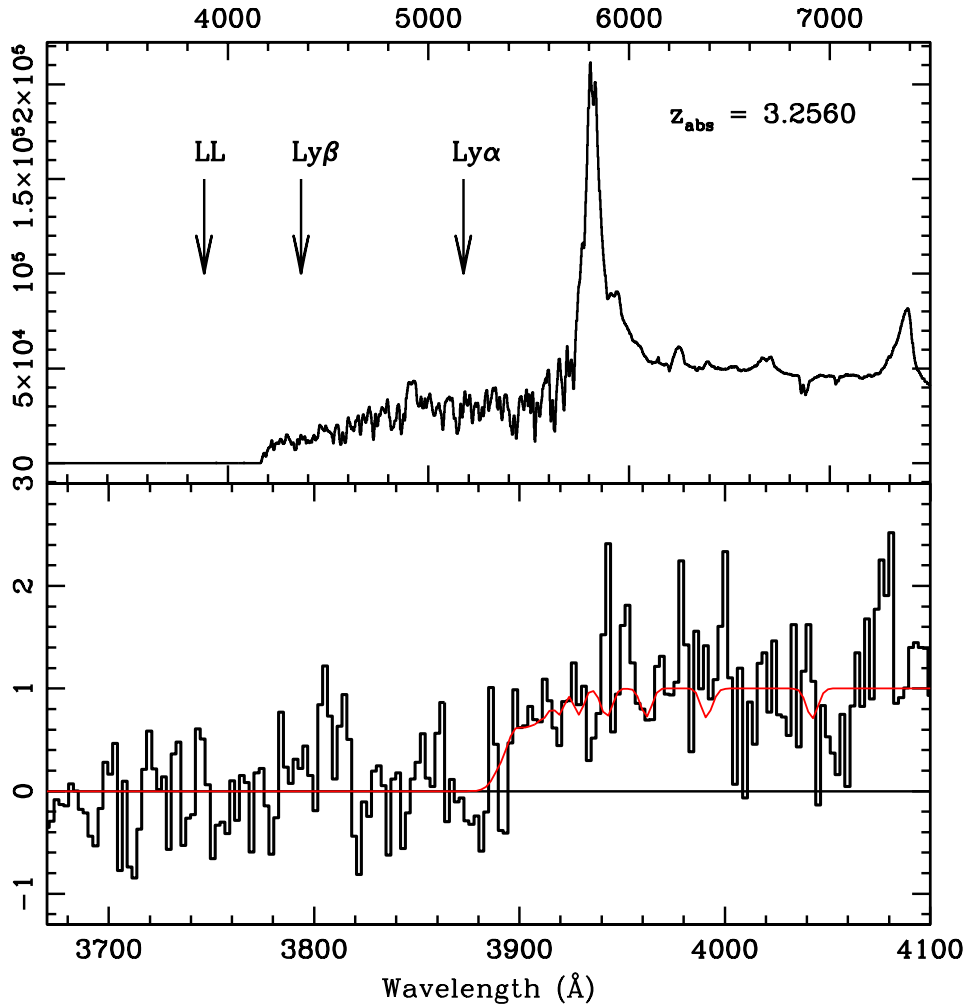}
\caption{\label{ll} The entire LRES spectrum (above) and an expanded
region of the spectrum showing the LL due to the $z=3.256$ absorber
(below). The fit corresponding to model (2d) (see section \ref{cont})
is shown in the expanded spectrum.}
\end{center}
\end{minipage}
\end{figure*}

Here we present an analysis of D/H in a newly discovered $z=3.256$
absorption system towards PKS1937-1009 (see Fig.~\ref{img}, Fig.~\ref{ll}).
Coincidentally, \citet*{Burles98} measured D/H in a
different absorber at $z=3.572$ towards this same QSO.  The new
absorber is not connected with the $z=3.572$ absorber, and we expect
it to give an independent measurement of D/H.

\section{Observations}

PKS1937-1009 was first identified as a quasar with $z_{em} = 3.787$ by
\citet{Lanzetta91}.  Our observations of PKS1937-1009 were taken on
Keck I with the HIRES and LRES instruments.  The LRES observations
were taken on 9th August, 1996 and the HIRES observations were taken
over the 31st May and 1st June, 1997.  The HIRES spectrum was taken
using the C5 decker, giving a resolution of $8.8$ \kms\ FWHM.  The
LRES spectra were taken using the $300$/$5000$ grating, giving a
resolution of $\sim 7.4$ \AA\ FWHM.

The data reduction was carried out using standard IRAF routines.  The
raw CCD frames were flat fielded and ADU counts corrected to photon
counts.  Medians were formed of the 2D images to eliminate most of the
cosmic rays.  An observation of a standard star was used to trace the
spectra in the 2D CCD image (for both LRES and HIRES data) and that
trace was adopted in carrying out the extraction from 2D to 1D
spectra.  The stellar trace was applied to the quasar exposures,
allowing it to translate in the spatial direction to account for
shifts between the relative placing of the quasar and star along the
spectrograph slit.  Optimal extraction was done, using the weights
derived from the standard star spectrum.  Error arrays were propagated
throughout the extraction procedure.  Polynomials were fitted to ThAr
lamp exposures to relate pixel number to wavelength. We checked for
shifts between calibration exposures bracketing the quasar
observations and time-interpolated where necessary.  The wavelength
scales were corrected to a heliocentric reference frame and all
spectra were rebinned onto a linear wavelength scale.

The LRES spectra were flux calibrated using the standard star
observations to recover the correct underlying spectral shape of the
quasar continuum.  No flux calibration was applied to the HIRES
spectra.  The final HIRES spectrum covers wavelengths from $4100 -
6400$ \AA\ with a signal to noise ratio (S/N) per pixel of $30$ at
$5000$ \AA.  The LRES spectrum covers the range $3700 - 7400$ \AA,
with a S/N per pixel of $\sim 200$ at $6000$ \AA.

\section{Analysis}

\begin{figure*}
\begin{minipage}{140mm}
\begin{center}
\includegraphics[width=1.0\textwidth]{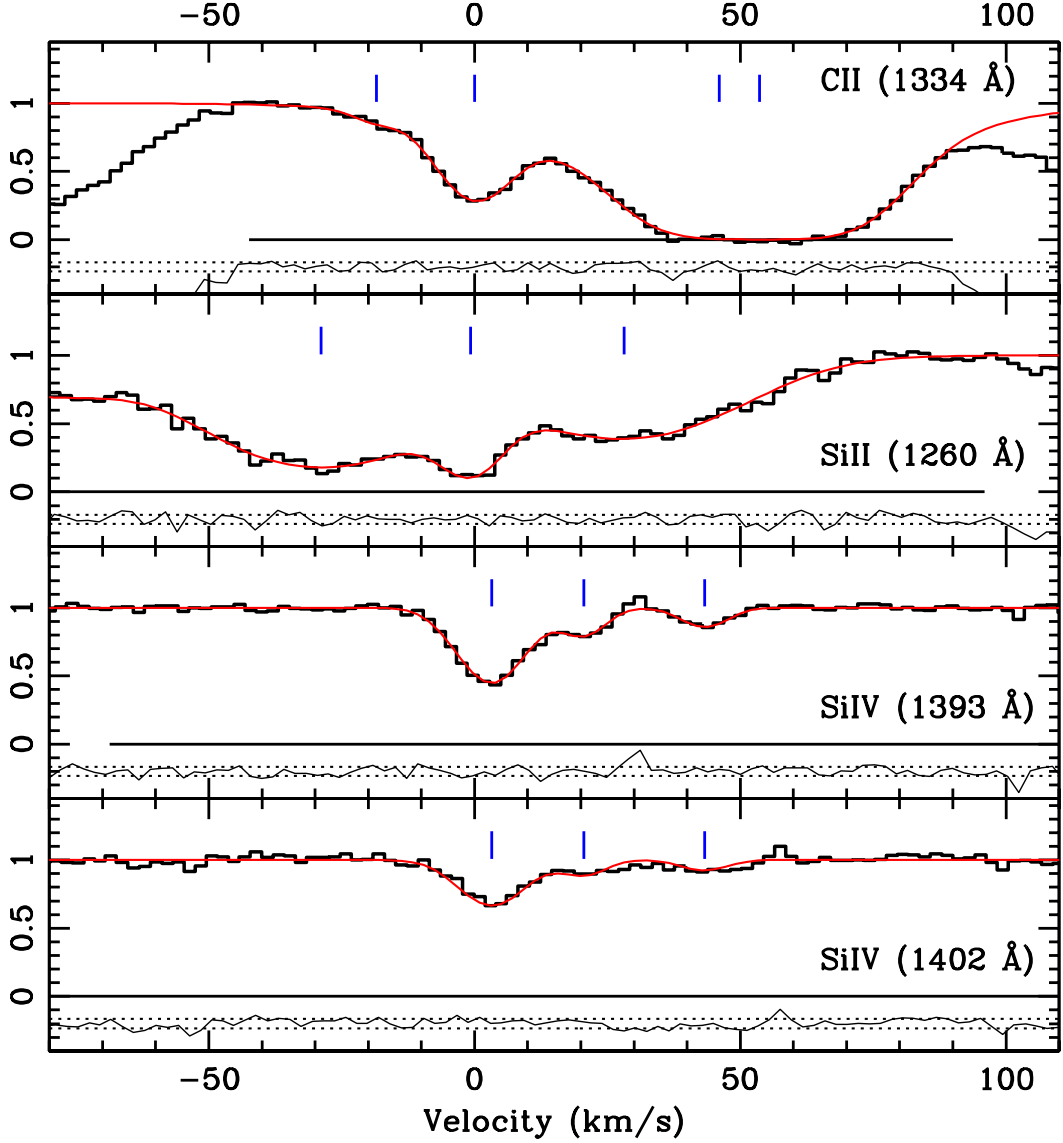}
\caption{\label{metals} The \carb2, \s2\ and \si4\ lines along with
their best fitting models used to derive the parameters found in table
\ref{metal}.  The velocity zero point is the redshift of the \carb2\
line, $z=3.256009$.  The histogram shows the data and the curves show
the fitted model.  The absorption lines away from the $v=0$ position
in the top two panels are due to \lya\ forest lines.  The three
components in the bottom two panels are all due to \si4.  The
normalised residuals ($({\rm data}-{\rm model})/{\rm error}$) and
their $1 \sigma$ ranges are shown below each spectrum, offset from the
zero level and scaled for clarity.  Horizontal lines at zero flux show
the fitting regions.}
\end{center}
\end{minipage}
\end{figure*}

\begin{figure*}
\begin{minipage}{140mm}
\begin{center}
\includegraphics[width=1.0\textwidth]{pics/simple.epsi}
\caption{\label{lya} The \lya\ and \lyb\ lines, along with a fit to
data.  The fit corresponds to model (2d), a two \hi\ component fit
where the level of the continuum is allowed to vary above both the
\lya\ and \lyb\ lines.  The continuum over the \lya\ line was
generated using a 3rd order chebyshev polynomial.  The velocity zero
point is the redshift of the \carb2\ line. The histogram shows the
data and the narrow curve shows the fitted model.  The normalised
residuals are shown beneath each graph.}
\end{center}
\end{minipage}
\end{figure*}

\begin{figure*}
\begin{minipage}{140mm}
\begin{center}
\includegraphics[width=1.0\textwidth]{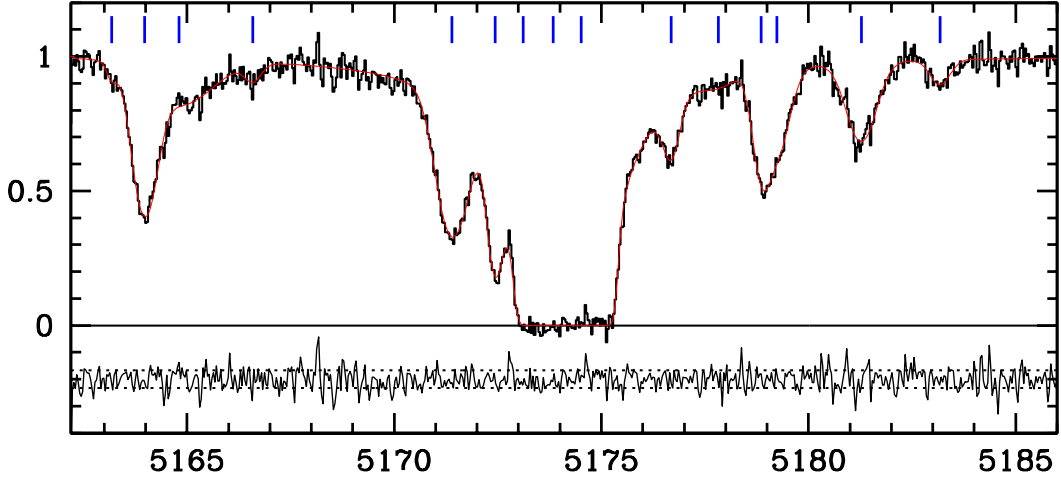}
\caption{\label{biglya} The \lya\ line showing the damping wings and
the best fitting model, model (2d).  The histogram shows the data and
the narrow curve shows the fitted model.  The normalised residuals are
shown below the spectrum.}
\end{center}
\end{minipage}
\end{figure*}

We fit Voigt profiles to the relevant absorption lines using the
program VPFIT\footnote{Carswell et al.,
http//www.ast.cam.ac.uk/$\sim$rfc/vpfit.html}.  This uses a $\chi^2$
minimisation procedure to find the best fitting redshift, column
density (\cm2) and $b$ parameter (by definition, $b=\sqrt{2}\sigma$
where $\sigma$ is the Gaussian velocity dispersion) for each
absorption line.  The diagonals of the covariance matrix give the
error estimates on each of the fitted parameters.  These errors assume
that the $\chi^2$ parameter space around the best fitting solution is
parabolic, which is not always the case. In particular, if there are
degeneracies in the fitted parameters, the $\chi^2$ parameter space
needs to be explicitly explored to find the correct error ranges.  For
the majority of the transitions we fit in the $z=3.256$ system, the
errors given by the covariance matrix are sufficiently accurate.
However, to find the errors on the best fitting value of D/H, we do
not use the errors given by VPFIT, but instead explicitly calculate
$\chi^2$ as a function of D/H, as described in section \ref{d2h}.

\subsection{Absorption lines present in the $z=3.256$ system}

\carb2 (1334), \s2 (1260) and \si4\ (1393 and 1402) (Fig.~\ref{metals}),
\hi\ \lya\ and \lyb\ (Fig.~\ref{lya}) transitions are present at
$z=3.256$ in the HIRES spectrum.  There is a narrow line in the blue
wing of the \hi\ \lya\ line which we believe is \di.  The $z=3.256$
Lyman limit (LL) is present in the LRES spectrum (Fig.~\ref{ll}).  All
higher order \hi\ transitions are at wavelengths shorter than the LL
due to the higher redshift absorber at $z=3.572$, and cannot be
detected in our HIRES spectrum.

The HIRES spectrum covers the position of \k3\ (1206.5) and \fe2\
(1144), but they are blended with \hi\ absorption in such a way that
no useful constraints are possible.

Both the \s2\ and \carb2\ lines fall in the \lya\ forest, where there is
the danger that an \hi\ forest line may be misidentified as a metal
line.  However, we are confident that we have correctly identified
\s2\ and \carb2\ for two reasons.  Firstly, their widths are much smaller
than expected for the average \lya\ forest \hi\ line.  Secondly, they
fall within a few \kms\ of the expected redshift based on $z$(\hi) and
$z$(\si4).  Unfortunately, both \s2\ and \carb2\ lines are to some extent
blended with \hi\ \lya\ absorption, which may be hiding weaker \s2\
and \carb2\ components.

The \si4\ lines are at wavelengths longer than the quasar \hi\ \lya\
emission and so are not blended with forest lines.

The \carb2\ and \s2\ appear to be at the same redshift, $z$(\carb2)$ =
3.256009 \pm 0.000006$ and $z$(\s2)$ = 3.256001 \pm 0.000009$.
 
There are three \si4\ components, one main component at $+3.4 \pm 0.4$
\kms (redward) relative to the \carb2/\s2\ position, and two lower column
density components at $+21.4 \pm 0.7$ \kms\ and $+43.4 \pm 0.7$ \kms.

The metal line parameters and their errors are given in Table \ref{metal}.

\begin{table*}
 \centering \begin{minipage}{140mm}
\begin{center} 
\caption{\label{metal} The absorption line parameters and their errors,
found using from the diagonals of the covariance matrix in VPFIT, for
\carb2, \s2\ and the three \si4\ components.}

\begin{tabular}{r l l l}

\hline 
Ion & \multicolumn{1}{c}  {log($N$) (\cm2)} & \multicolumn{1}{c} %
z & \multicolumn{1}{c} {$b$ (\kms)} \\
\hline 
 \carb2  & 13.600 $\pm$ 0.045 & 3.256009 $\pm$ 0.000006 & 6.9 $\pm$ 0.7 \\
 \s2 & 12.875 $\pm$ 0.093 & 3.256001 $\pm$ 0.000009 & 5.0 $\pm$ 1.4 \\
 \si4 & 12.820 $\pm$ 0.011 & 3.256057 $\pm$ 0.000002 & 6.5 $\pm$ 0.3 \\
 \si4 & 12.138 $\pm$ 0.039 & 3.256303 $\pm$ 0.000005  & 1.3 $\pm$ 1.3  \\
 \si4 & 11.970 $\pm$ 0.036 & 3.256626 $\pm$ 0.000006  & 2.8 $\pm$ 1.3  \\
\hline 
\end{tabular}
\end{center}
\end{minipage}
\end{table*}

\subsection{\hi\ column density}
The $z=3.256$ Lyman limit is optically thick, so there is a lower
limit on the total log~$N$(\hi) of $\sim 17.6$.  There are also
damping wings at the \lya\ line, which further constrain $N$(\hi)
(Fig.~\ref{biglya}).  There might be another explanation for these
damping wings; perhaps the continuum dips over the \lya\ line, due to
an emission line on either side of the \lya\ line.  However, composite
QSO spectra [e.g. \citet{VandenBerk01}] display no emission lines
associated with the quasar redshift at the relevant wavelengths to
cause this apparent dip in the continuum.  We believe our two
estimates of the continuum, described below, adequately account for
any uncertainty in the shape of the continuum.

\subsection{Continuum placement}
\label{cont}

To generate the continuum we fitted regions that were apparently free
from absorption around each absorption feature, with low order
Chebyshev polynomials.  The damping wings present in the \hi\ \lya\
line provide the best constraint on the total $N$(\hi), but the
derived $N$(\hi) value is sensitive to the placement and shape of the
continuum above the damping wings.  To take this into account, we
fitted two different continua above the \lya\ line using 2nd and 3rd
order Chebyshev polynomials.  We also allow the continuum to vary
above the \lya\ and \lyb\ lines by multiplying it by a parameter,
initially set to $1$, which is allowed to vary during the $\chi^2$
minimisation.  We use two independent parameters to vary the continuum
level above the \lya\ and \lyb\ lines.  An example of the typical
fitted values and errors for the two continuum parameters is given in
Table~\ref{param}.  For all our models, the fitted continua are within
$5$\% of our initial guess for the continuum.

\subsection{\hi\ velocity structure}

We try to determine the velocity structure in two ways: firstly, by
fitting the available \hi\ transitions with as few components as
needed to obtain an acceptable fit, and secondly, by looking at the
metal lines to indicate the possible position of \hi\
components. Ideally we would like to fit many higher order \hi\ Lyman
transitions to help determine the velocity structure, but only the
\lya\ and \lyb\ lines are available.

A single \hi\ component cannot satisfactorily fit both the \lya\ and
\lyb\ lines, so at least two components are present.
  
We expect at least four hydrogen components, one at $z$(\carb2/\s2) and three
corresponding to the \si4\ component positions.  However, we don't
know what \hi\ column density is associated with each one.  We expect
the \carb2\ and \s2\ components to have a greater proportion of their
associated H gas in the form of \hi, as their ionization potentials
are close to that of \hi.  Therefore, if there is no significant
metallicity variation across the absorption complex, an \hi\ component
near $z$(\carb2/\s2) is likely to be the dominant component in the \lya\ and
\lyb\ line.

Below we describe the velocity models we explored.

\subsubsection{Model (1): two \hi\ components, $z$(\hi) tied to $z$(\carb2)}
 We initially assemble a model using two \hi\ components, with the
redshift of the bluer component tied to the \carb2\ redshift. In this fit
we include \di\ components for both \hi\ components.  We tie the
redshifts of corresponding \di\ and \hi\ components together, and fit
a single D/H to both components (we discuss this assumption in section
\ref{fixd2h}).  Since the redder \di\ component is heavily blended
with the bluer \hi\ component, we tie its $b$ parameter to the $b$
parameter of the corresponding \hi\ component, assuming purely thermal
broadening (this assumption has no effect on the final D/H value; see
section \ref{bpar}).  We fit the summed total of the \hi\ column
density of both components, instead of individually fitting the column
densities of each component.  We do this because while the column
density in each component is not well constrained, the total
log~$N$(\hi) is.  This model does not fit the \di\ line in the \lya\
line adequately.  We can obtain an acceptable fit if we include very
weak absorption from some unknown contaminating metal line near the
\di\ position.  However, this results in D/H $< 0.6 \times 10^{-5}$
($1 \sigma$ upper limit).  This is smaller than even the lowest
estimates of the interstellar medium D/H, $\sim 1.0 \times 10^{-5}$
\citep{Hebrard03}.

\subsubsection{Model (2): two \hi\ components, $z$(\hi) not tied to $z$(metals)}
It is possible that the main \hi\ component does not fall at exactly
the same position as the \carb2\ and \s2\ lines.  This could be due to
several reasons: perhaps an error in the wavelength calibration of a
few \kms, or a gradient in the metallicity across the cloud.  For this
reason, we explore a model that does not tie the redshifts of the \hi\
components to that of the metal lines.  This model is identical to the
model above, except it no longer ties the redshift of the bluer \hi\
component to the \carb2\ line.  This model gives an acceptable fit to
the data.  The position of the bluer hydrogen component is $-3.5 \pm
0.7$ \kms\ relative to \carb2.  The redder \hi\ component is at $+36
\pm 4$ \kms.

\subsubsection{Model (3): four \hi\ components, $z$(\hi) tied to $z$(metals)}

We included four \hi\ components, the redshift of each tied to a metal
line component redshift (one corresponding to the \carb2/\s2\ position
and one for each \si4\ component).  We include \di\ for all components
where log~$N$(\hi)$ > 17.0$.  This does not give a satisfactory fit to
the data: the narrow \di\ line in the \lya\ line in particular is very
poorly fit. As in model (1), the fit could be improved by including
very weak contaminating absorption near the \di\ position.  Again,
this gives a low D/H, $< 0.5 \times 10^{-5}$, much lower than the ISM
D/H.

Model (2) gives the only acceptable fit to the data.  Models (1) and
(3) can give an acceptable fit, but we need to include an unidentified
metal line blended with \di, and there is no other evidence that this
is present.  If we do include such a line, the measured D/H in this
absorber is much smaller than the ISM D/H.

We split model (2) into four sub-models, using (2a) a 2nd order
polynomial fixed continuum, (2b) a 3rd order polynomial fixed
continuum, (2c) a 2nd order polynomial varying continuum and (2d) a
3rd order polynomial varying continuum.

\begin{table*}
 \centering \begin{minipage}{140mm}
\begin{center} 
\caption{\label{param} The best fitting parameters of the lines used
to fit the \lya\ and \lyb\ lines and the LL of the $z=3.256$ absorber.
The errors are those returned by VPFIT, calculated using the diagonals
of the covariance matrix.  The first two \hi\ components in the table
have associated \di\ lines which we included.  We fitted the total
column density of both these components - the column density entry
next to the first \hi\ component give the value of and error on the
total column density of both \hi\ components.  The bottom two rows
give the best fitting continuum parameters (which are multiplied by
the initial guess for the continuum level) for the \lya\ and \lyb\
regions.}

\begin{tabular}{r l l l}
\hline 
Ion & \multicolumn{1}{c}  {log$N$ (\cm2)} & \multicolumn{1}{c} %
z & \multicolumn{1}{c} {$b$ (\kms)} \\
\hline 
   \hi   &    18.255  $\pm$  0.018 (total) & 	3.255959  $\pm$  0.000008 &  15.8 $\pm$ 0.3\\	
   \hi   &   & 	3.256515  $\pm$  0.000059 &  15.6 $\pm$  1.3 \\ 
   \di   &    13.393  $\pm$  0.069 (total) & 	3.255959 (tied to \hi)  &  11.6 $\pm$  0.9\\
   \di   &   & 	3.256515 (tied to \hi) &  11.1 (tied to \hi) \\ 
   \hi   &    13.515  $\pm$  0.014 &  3.253948  $\pm$  0.000007 &  26.9 $\pm$   0.8\\  
   \hi   &    12.775  $\pm$  0.030 & 	3.258296  $\pm$  0.000011 &  15.1 $\pm$   1.3\\
   \hi   &    13.73  $\pm$  0.18 & 	2.589492  $\pm$  0.000042 &  15.4 $\pm$   3.3\\
   \hi   &    14.12  $\pm$  0.083 & 	2.588996  $\pm$  0.000059 &  33.9 $\pm$   3.4\\
   \hi   &    12.09  $\pm$  0.17 & 	3.259237  $\pm$  0.000112 &  28.4 $\pm$  14.0\\
   \hi   &    12.97  $\pm$  0.34 & 	3.260086  $\pm$  0.000042 &  15.5 $\pm$   3.3\\
   \hi   &    13.09  $\pm$  0.26 & 	3.260394  $\pm$  0.000146 &  23.4 $\pm$   6.4\\
   \hi   &    13.118  $\pm$  0.014 & 	3.262072  $\pm$  0.000010 &  27.3 $\pm$   1.0\\
   \hi   &    12.399  $\pm$  0.054 & 	3.263637  $\pm$  0.000025 &  18.4 $\pm$   2.7\\
   \hi   &    13.362  $\pm$  0.036 & 	3.247846  $\pm$  0.000006 &  21.5 $\pm$   1.0\\
   \hi   &    13.195  $\pm$  0.073 & 	3.248529  $\pm$  0.000114 &  63.0 $\pm$   8.7\\
   \hi   &    11.89  $\pm$  0.23 & 	3.247192  $\pm$  0.000030 &   7.7 $\pm$   4.6\\
   \hi   &    12.15  $\pm$  0.11 & 	3.249996  $\pm$  0.000035 &  14.8 $\pm$   4.1\\
   \hi   &    14.179  $\pm$  0.031 & 	2.593103  $\pm$  0.000016 &  32.6 $\pm$   3.2\\
   \hi   &    12.63  $\pm$  0.38 & 	2.592337  $\pm$  0.000129 &  20.3 $\pm$  13.8\\
   \hi   &    13.21  $\pm$  0.11 & 	2.593948  $\pm$  0.000064 &  27.9 $\pm$   5.9\\
   \hi   &    13.093  $\pm$  0.094 & 	2.587331  $\pm$  0.000015 &  12.1 $\pm$   2.5\\
   \hi   &    13.548  $\pm$  0.038 & 	2.587413  $\pm$  0.000031 &  53.4 $\pm$   5.5\\
   \hi   &    12.53  $\pm$  0.10 & 	2.586028  $\pm$  0.000024 &  11.2 $\pm$   3.4\\
   cont(\lya)   &     1.0035  $\pm$  0.0027 & &   \\ 
   cont(\lyb)    &     0.983  $\pm$  0.014 & &   \\ 
\hline 
\end{tabular}
\end{center}
\end{minipage}
\end{table*}

\subsubsection{Two unresolved questions about the best-fitting model}

The best fitting model seems to fall below the data at $-90$ \kms\ in
the \lyb\ line (Fig.~\ref{lya}).  There could be several reasons for
this. It is possible that the apparent discrepancy is due to noise, or
there may be a weak, narrow \lya\ forest line near $-75$ \kms.
However, such a line would need to have a width much smaller than the
median width of forest lines.  A metal line could have a narrow width,
but the higher redshift system at $z=3.572$ does not have any metal
lines which fall at this position. No other absorption systems were
identified that could explain a metal line appearing at this position.
Instead there may be a weak cosmic ray at this position that has not
been completely removed.  Two of the three 2D raw images combined to
form the final spectrum show cosmic rays at the positions
corresponding to $-75$ to $-95$ \kms\ in the \lyb\ line.  We
tentatively conclude that a noise spike is the cause of this
discrepancy.  We checked whether this discrepancy has a significant
effect on the measured D/H by fitting the \lya, \lyb\ lines and LL
excluding the region from $-75$ to $-95$ \kms.  We found that the best
fitting D/H value and its error are almost unchanged, as the tightest
constraint on $N$(\di) comes from the \lya\ \di\ line.
\\
\\
We would expect the \s2\ and \carb2\ redshifts to be consistent with
the redshift of the \hi\ line associated with them.  Instead we find
that the \s2/\carb2 position differs from the \hi\ position by $3.5
\pm 0.7$ \kms, or about $1.5$ pixels.  It is possible that there are
problems with the wavelength calibration.  This is unlikely, however,
given that the \s2\ and \carb2\ positions agree to within $1$\kms,
and that the typical wavelength accuracy of a polynomial fit to an arc
exposure from HIRES is better than $0.5$ \kms.  It is more likely that
our velocity model is wrong in some way.  The true velocity structure
of the absorption complex will almost certainly be more complicated
than our simple two component model.  However, we note for all other
velocity models that we considered, D/H was substantially {\it lower}
($< 0.6 \times 10^{-5}$) than the value that we quote for Model 2(b).
\\ 
\\
Both of these questions may be resolved by much higher S/N
observations of this absorption system.

\subsection{Ionization and metallicity}
\label{ionize}

\begin{figure*}
\begin{center}
\begin{minipage}{140mm}
\begin{center}
\includegraphics[width=1.0\textwidth]{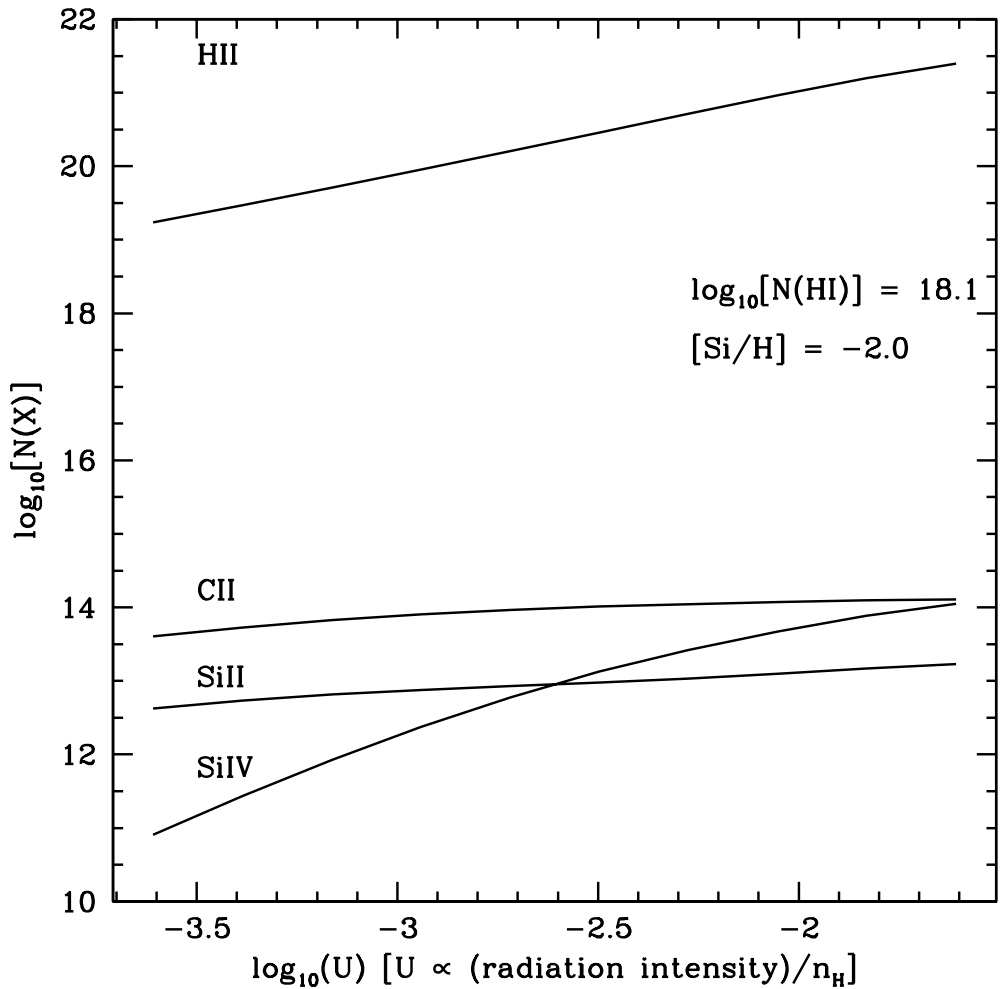}
\caption{\label{cloudy} The ionization model generated using CLOUDY.  The parameter $U$ is defined in section \ref{ionize}.}
\end{center}
\end{minipage}
\end{center}
\end{figure*}

We have $N$(\s2) and $N$(\carb2) in a single gas cloud, and $N$(\si4) in
three other clouds.  We use the program CLOUDY \citep{Ferland97} to
generate an ionization model for the absorption system.  When
generating the model, we assume that the bluest \si4\ component, the
\s2\ and \carb2\ component, and the bluest \hi\ component are all
associated with the same gas, even though they span $\sim 6$ \kms.  We
approximate the geometry of the cloud as a plane parallel slab
illuminated by a distant point source and assume the proportion of
metals to be solar.  For calculating ionization fractions, the
hydrogen particle density, $n_{H}$ and the incident radiation flux on
the cloud, $J_{\nu}$ are degenerate.  Instead of specifying either
$n_{H}$ or $J_{\nu}$, we use the parameter $U$:
$$ 
U = 2 \times 10^{-5} \frac{J_{912}/(10^{-21.5} \textrm{erg cm$^{-2}$ s$^{-1}$ sr$^{-1}$})}{n_{H}/(1 \rm{ cm^{-3}})}
$$
Here $J_{912}$ is the incident radiation at $912$ \AA.  We generate
ionization models for a range of metallicities and $U$ values, using a
Haardt-Madau ionizing radiation continuum at a redshift of $3.25$
\citep{HaardtMadau96}.  We have not explored the effect a different
ionizing radiation continuum may have on the ionization fractions.
$N$(\s2) and $N$(\si4) suggest that the system has a total metallicity
[Si/H] $\sim -2.0$. $N$(\carb2) is lower than expected for this [Si/H]
which might indicate that C is under abundant compared to Si (see
Fig.~\ref{cloudy}).  This is unlikely to be due to differential dust
depletion, since C has a lower condensation temperature than Si. We
would expect Si, not C, to be under abundant if there is any depletion
on to dust \citep{Lu96}.  However, an under abundance of C is
consistent with metallicity measurements in metal poor stars and \hii\
regions, which find [C/O] $\sim -0.5$ \citep{Akerman04}.  Since both
Si and O are $\alpha$-capture elements, it is reasonable to also
expect [C/Si] $\sim -0.5$ when [Si/H] $= -2$ (Max Pettini, 2004,
private communication).

With so few transitions we can only make broad generalisations about
conditions in the cloud.  We take a value of [Si/H] $-2.0$ with an
error of $\sim 0.5$ dex, where this error is dominated by the
uncertainty on $N$(\hi) in the bluest \hi\ component.

\subsection{{\em b} parameter test}
\label{bpar}

\begin{figure}
\begin{center}
\includegraphics[width=0.49\textwidth]{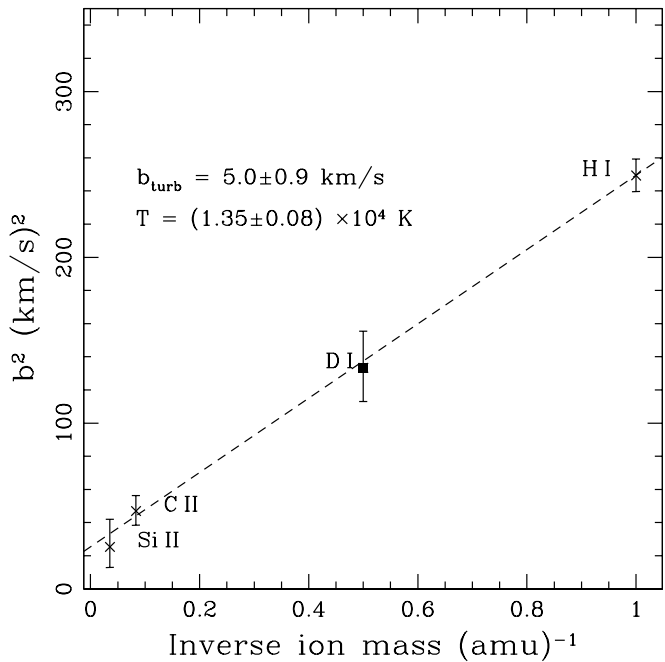}
\caption{\label{btest} The $b$ parameter squared (\kms)$^2$ versus the
inverse ion mass (amu$^{-1}$).  The dashed line is the least squares
line of best fit to the \s2, \carb2, and \hi\ points. \di\ is shown for
comparison.  The error bars in each case are calculated from the
1$\sigma$ errors given by VPFIT.}

\end{center}
\end{figure}

We can predict the $b$ parameter of the blue \di\ component from the
$b$ parameters of the \s2, \carb2 and \hi\ lines.  This is done by
measuring the thermal line broadening, $b_{\rm{therm}}$, and turbulent
broadening, $b_{\rm{turb}}$, by using the $b$ parameters from all
available ions.  If we assume the absorbing cloud has a thermal
Maxwell-Boltzmann distribution and any turbulence can be described by
a Gaussian velocity distribution, then the $b$ parameter for a
particular ion will be given by

\begin{equation}
b_{\rm{ion}}^2 = b_{\rm{therm}}^2 + b_{\rm{turb}}^2.
\end{equation}

Here $b_{\rm{therm}}^2 = \frac{2kT}{m}$, where $T$ is the temperature
of the gas cloud, $m$ is the mass of the absorbing ion and $k$ is
Boltzmann's constant.  $b_{\rm{turb}}^2$ represents the Gaussian
broadening due to small scale turbulence, and is the same for all
ionic species.  We plot $b^2$ against inverse ion mass in
Fig. \ref{btest}.  Subject to the assumptions above, all the ions in
the same cloud velocity space should lie on a straight line whose
intercept gives $b_{\rm{turb}}$ and slope gives the cloud temperature.
\carb2, \s2 and \hi\ all have relatively low ionization potentials
(IP); 24.4, 16.3 and 13.6 eV, respectively. There is evidence from
damped \lya\ absorbers that even `intermediate' IP ions, such as
Al\,{\sc iii} (IP 28.4 eV), have a similar velocity structure to lower
IP ions \citep{Wolfe00a}.  Thus we assume that the \s2\ and \carb2\
lines arise in the same gas as the \hi\, and fit a least squares line
of best fit to these three points.  This is shown as the dashed line
in Fig.~\ref{btest}. The \di\ point is plotted with its $1 \sigma$
error bars as given by VPFIT.  We find a temperature of $1.35 \pm 0.08
\times 10^4$K and a turbulent broadening of $5.0 \pm 0.8$\kms.  These
values suggest a $b$ parameter for the blue \di\ component of $11.7$
\kms.  This is consistent with our fitted $b$(\di) of $11.6 \pm 0.9$
\kms\ for the blue \di\ component.

There is no way to measure the $b$ parameter of the red \di\
component, since it is heavily blended with the \hi\ blue component.
We also cannot predict what $b$(\di) will be using the method above,
since we do not have any low ionization metal lines corresponding to
the red \hi\ component.  In our velocity models we explored the
maximum and minimum $b$(\di) values, corresponding to purely thermal
[$b$(\di)$= 1/\sqrt{2}\ b$(\hi)] and purely turbulent [$b$(\di)\ $=
b$(\hi)] broadening.  The measured D/H is not affected by which
broadening we assume.  Thus we arbitrarily choose pure thermal
broadening, assuming that the the conditions in the red component are
similar to the blue component.

\subsection{Are we seeing deuterium?}
\label{isd}
We expect we are seeing D, for three reasons.  The line we
have identified as \di\ occurs near the expected position for \di,
given the position of the main \hi\ absoprtion and the position of the
\s2\ and \carb2\ lines.  The $b$ parameter of the \di\ line is consistent
with that expected for \di\ based on the fitted $b$ parameters of the
\hi\ component and the \s2\ and \carb2\ lines.  Finally, the line has a
width substantially narrower than the median for forest lines, which
means it is unlikely to be an \hi\ \lya\ forest line.  A metal line
from another absorption system could have a line width this narrow, but
we know of no other absorption systems towards this quasar which have
metal lines that fall at the position of the \di\ line.  These
arguments all apply only to the bluer \di\ component.  We cannot
independently measure the position, $b$ parameter or column density of
the the redder \di\ component.  However, we assume that D/H is the
same in both components for the reasons given in section \ref{fixd2h}.

\subsection{The validity of our assumption of constant D/H across different components}
\label{fixd2h}
We assume in our models that D/H is the same for all components
showing \di.  D is destroyed in star formation, so if the components
have different metallicities, it is conceivable that they will also
have different D/H values.  Unfortunately, we cannot directly measure
the metallicity of individual components due to the \lya\ forest
contamination around the \s2\ and \carb2\ lines.  However, we believe
that even if the metallicity is significantly different in different
components, D/H will still be very similar.  Even in systems where the
metallicity has been found to vary considerable between components in
the same absorption system \citep[e.g.][]{Kirkman03,Burles98}, the
difference is rarely more than $\sim 1.0$ dex.  Theoretical models
\citep{Fields96,Prantzos01} predict that D/H remains roughly constant
with metallicity until metallicities approaching $1/5$ solar are
reached.  Since [Si/H] $\sim -2.0$ for this entire absorption complex
(see section \ref{ionize}), we expect D/H in this absorber to be very
close to the primordial value.  Even if there are components with
[Si/H] as high as $\sim -1.0$ present, we don't expect their D/H to be
significantly lower than that of any lower metallicity components.

\subsection{D/H in the {\em z}=3.256 absorber}

\label{d2h}
\begin{figure*}
\begin{center}
\begin{minipage}{0.9\textwidth}
	\includegraphics[width=0.49\textwidth]{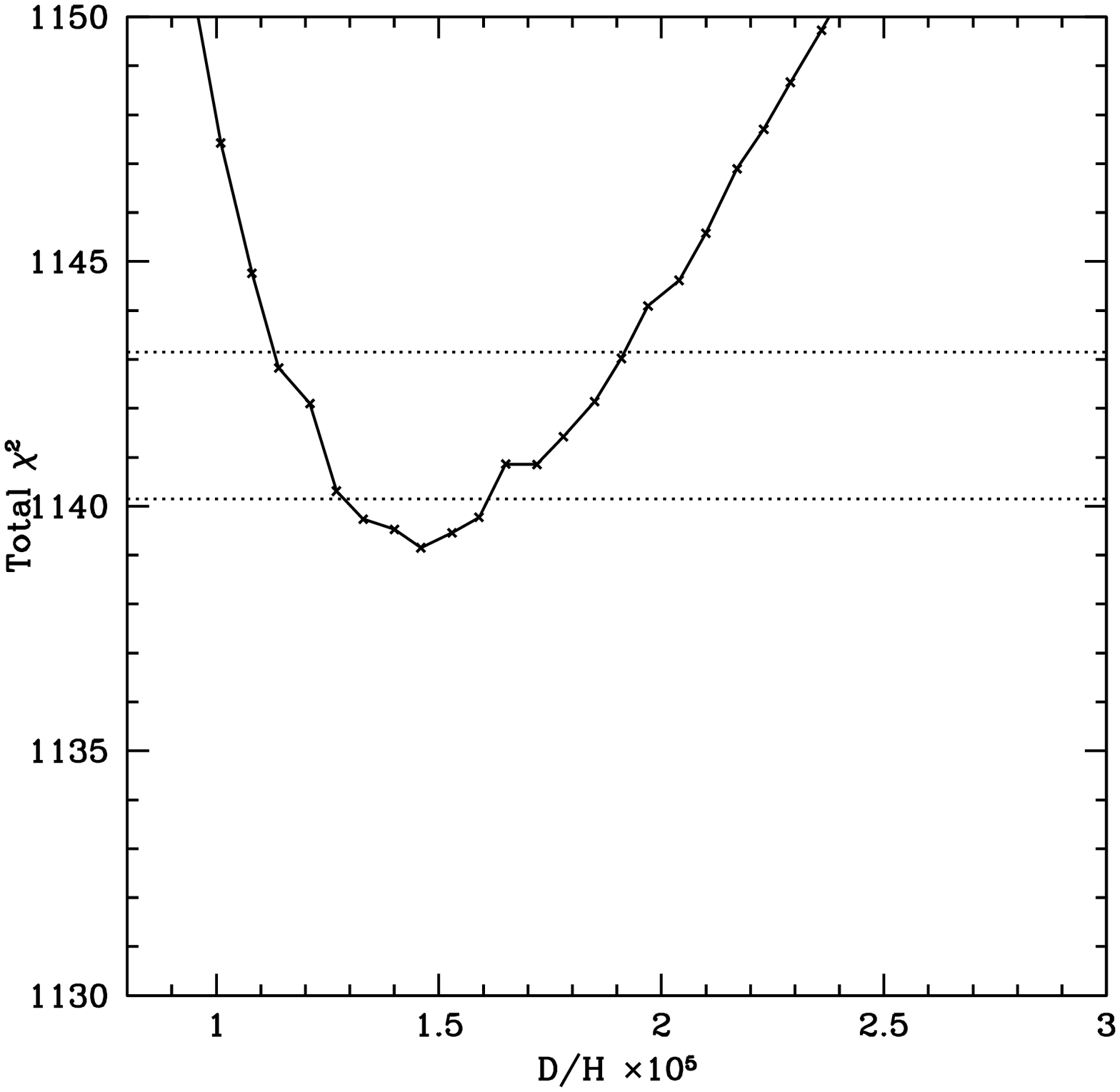}
	 \includegraphics[width=0.49\textwidth]{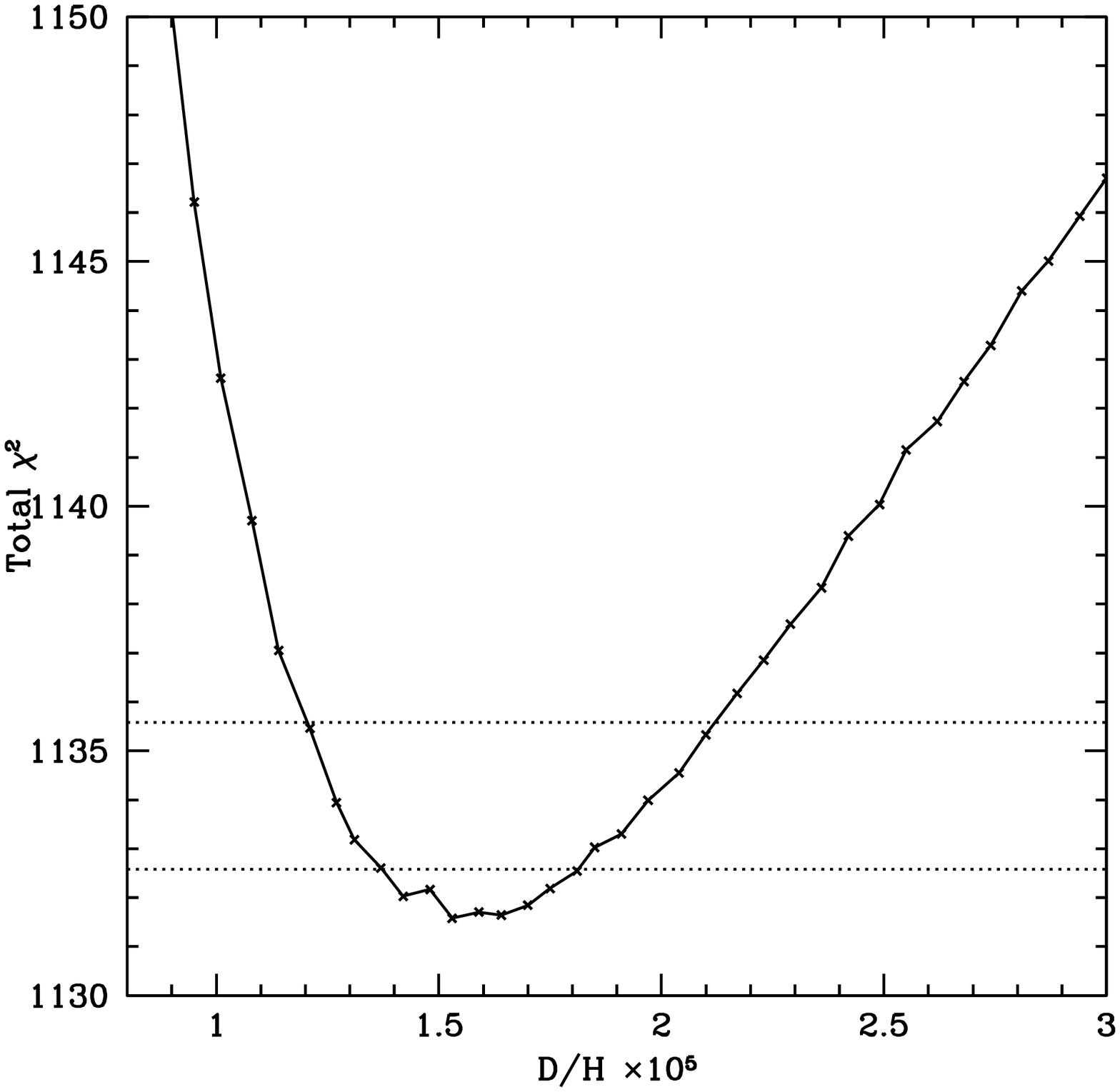}
\end{minipage}
\begin{minipage}{0.9\textwidth}
	 \includegraphics[width=0.49\textwidth]{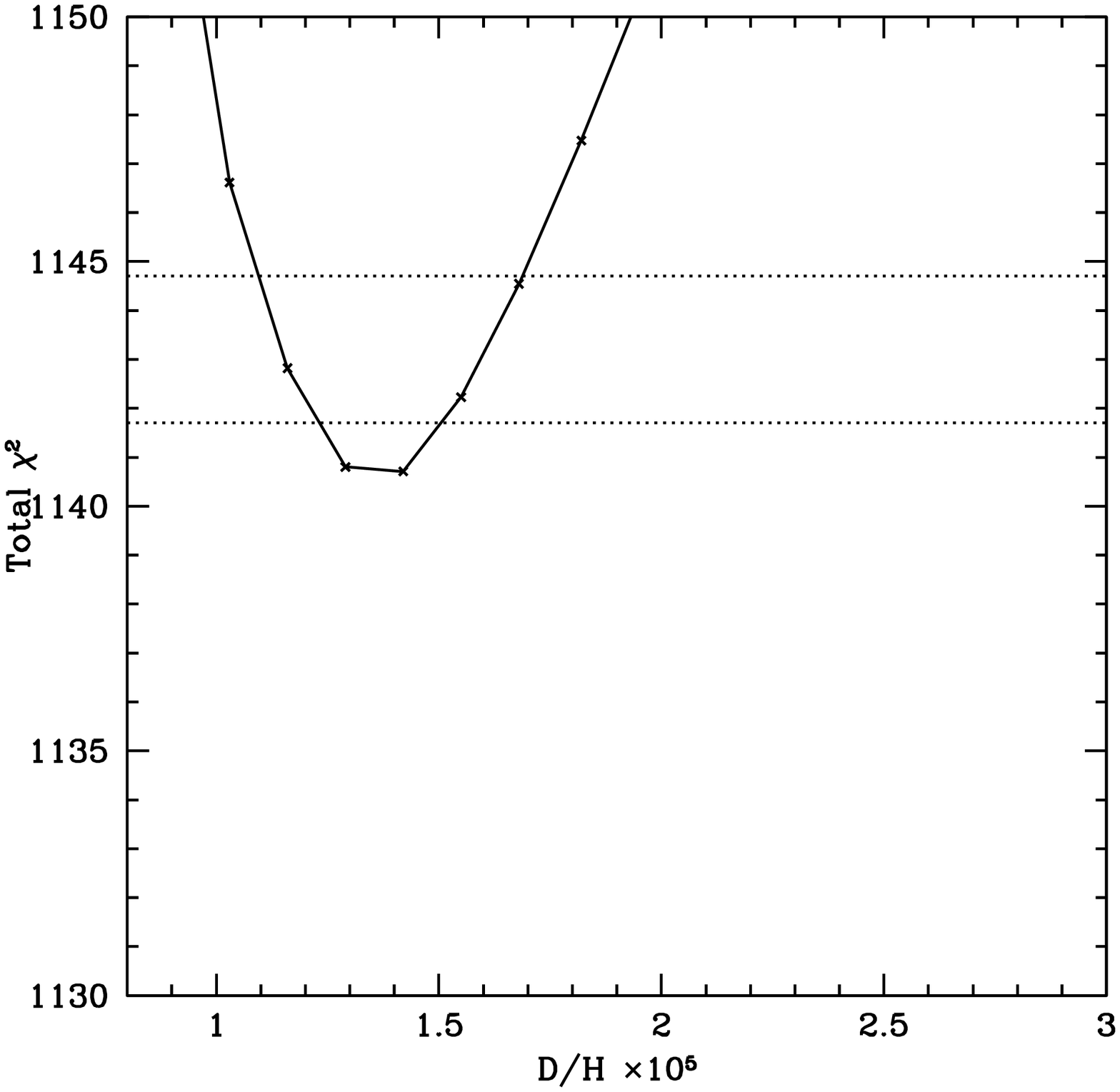}
	 \includegraphics[width=0.49\textwidth]{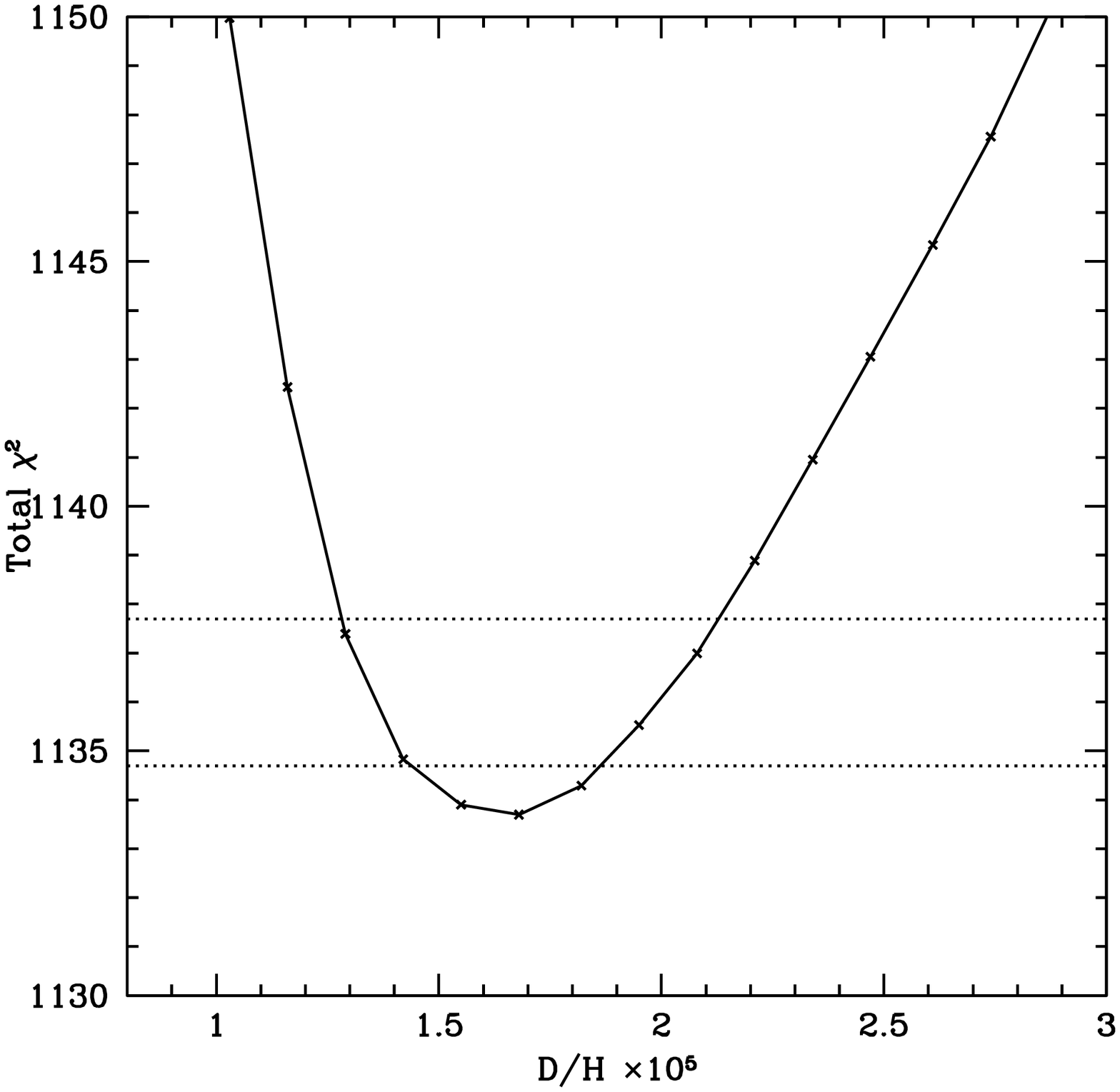}
\end{minipage}
\caption{\label{graph} $\chi^2$ vs D/H plots.  The horizontal lines
show the $68.3$\% and $95.4$\% confidence levels.  The upper graphs
are for models (2c) and (2d), where the continuum level is allowed to
vary during the fit.  The lower models are (2a) and (2b), where the
continuum levels are fixed.  The plots on the left use models with a
2nd order continuum fit over the \lya\ line, the plots on the right
use a 3rd order continuum.}
\end{center}
\end{figure*}

We have the values of $N$(\hi) and $N$(\di) and their errors for
models (2a) - (2d), but we prefer to use an alternate method to find
D/H in this absorber.  Only two small regions constrain $N$(\di), and
the errors on $N$(\di) are relatively large.  As we saw in
\citet{Crighton03}, when the errors given by VPFIT are large, it is
often a sign that the $\chi^2$ parameter space around the best fitting
solution is not symmetrical.  In this case the $\chi^2$ space should
be explored explicitly to find the correct error ranges.

We have generated a list of $\chi^2$ values as a function of D/H.  For
each value of D/H, we fix D/H and fit the \lya, \lyb\ lines and the
LL, varying all other parameters to minimise $\chi^2$.  In this case
we do not fit the the total $N$(\hi) and $N$(\di), since we are not
interested in the errors on individual parameters, just the total
$\chi^2$ value.  In this fit we tie the redshifts of each \di\ line to
its corresponding \hi\ line.  We also tie the $b$ parameter of the red
\di\ component to its correponding \hi\ component, assuming thermal
broadening, as discussed in section \ref{bpar}. 

We calculate one of these graphs for an appropriate range of D/H
values for each model, (2a) - (2d) (Fig.~\ref{graph}).  The
distribution of $\Delta\chi^2 \equiv \chi^2 - \chi^2_{\rm{min}}$ for
each graph is the same as that of a $\chi^2$ distribution with a
number of degrees of freedom equal to the number of fixed parameters
(in this case, one - D/H). Here $\chi^2_{\rm{min}}$ is the smallest
value of $\chi^2$ for a particular graph.  For each point on each
graph, we are fitting $990$ data points with $\sim 47$ parameters,
giving $\sim 943$ degrees of freedom.  The typical reduced $\chi^2$
for our best fitting models is $1.2$.

D/H is more tightly constrained in the cases without a floating
continuum level, which is expected as there are fewer free parameters.
The 3rd order continuum allows a better fit to the data than the 2nd,
and is consistent with a larger D/H value. The predicted $68.3$\%
range for D/H for the 3rd order continuum models (2b and 2d) is
$1.6^{+0.25}_{-0.30} \times 10^{-5}$, with a $95.4$\% range of
$1.6^{+0.5}_{-0.4} \times 10^{-5}$.  The D/H ranges for the 2nd order
continuum models are lower, $\sim 1.45 \pm 0.2 \times 10^{-5}$.  Model
(2d), which takes the uncertainty in the continuum level [and any
effect that may have on the uncertainty in the $N$(\hi) value] into
account, gives the best estimate of D/H in this absorber.  It is worth
noting that these error ranges are consistent with the ranges for D/H
that would be derived by using the VPFIT best fitting values and
errors given in Table.~\ref{param}.

\section{Comparison with D/H measured in other QSO absorbers}
\label{compd}

\begin{figure}
\begin{center}
\includegraphics[width=0.49\textwidth]{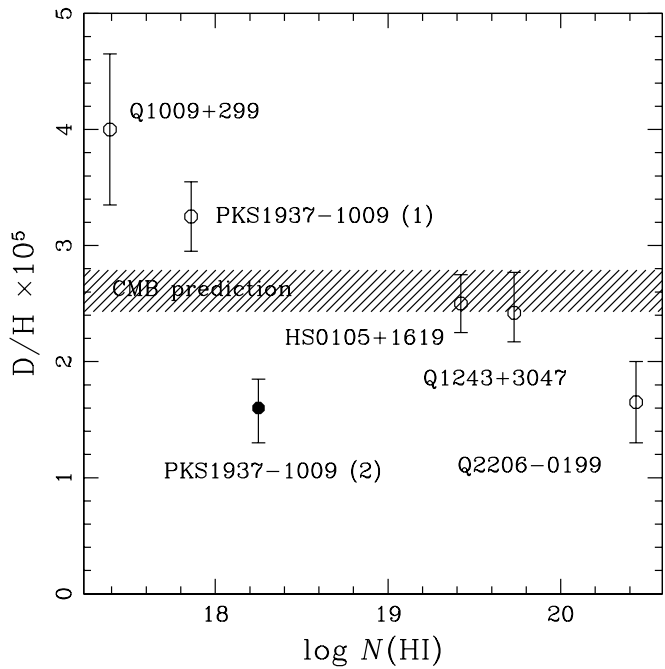}
\caption{\label{d2hvN} D/H measurements as a function of log$N$(\hi)
in the absorber where the measurement was made.  The filled circle is
the D/H value from this paper.  The hashed region is the CMB +
standard BBN prediction for D/H from \citet{Coc04}.  References for
the other D/H measurements are given in section \ref{compd}.}

\end{center}
\end{figure}

In this section we compare other D/H detections with our own
detection, and the predicted D/H value from CMB and standard BBN.
There are many upper limits on D/H in QSO absorbers that are much
higher than the CMB predicted D/H. Examples of such systems are
towards Q0014+813 \citep{Carswell94, Rugers96, Rugers96b, Burles99},
Q0420-388 \citep{Carswell96}, BR 1202-0725 \citep{Wampler96a}, PG
1718-4801 \citep{Webb97a, Tytler99, Kirkman01, Crighton03} and
Q0130-4021 \citep{Kirkman00}.  We do not include these in our
comparison, since they are clearly consistent with the predicted D/H
value.

We now describe six D/H detections in QSO absorbers, five of which we
include in our comparison.  We exclude one of the detections, towards
Q0347-3819, for the reasons outlined below in section \ref{Lev}.

\subsection{$z=3.572$ towards PKS1937-1009} 

\citet{Burles98} have presented the most comprehensive analysis of
this system.  Absorption attributed to \di\ is observed in two
transitions.  The measured $b$(\di) matches the predicted $b$ found
using the \s2, \carb2\ and \hi\ $b$ parameters.  This is a grey LL
system. $N$(\hi) was measured using the drop in flux at the LL and
initially estimated by \citet{Tytler_nat_96} as log $N$(\hi) $=17.94 \pm 0.05$.
This was revised in \citet{Burles97} to $17.86 \pm 0.02$ after
\citet{Wampler96b} and \citet{Songaila97} pointed out that the
continuum level of Tytler et al. was too high.  The final D/H value
from \citet{Burles98} is $(3.3 \pm 0.3) \times 10^{-5}$.

\subsection{$z=2.504$ towards 1009+2956} 

This system was analysed by \citet{Burles98b}.  It has a grey LL and
log $N$(\hi) $=17.39 \pm 0.06$, measured using the drop in flux at the
LL.  \di\ is seen in only a single transition. The authors believe it
is likely some contamination is present, so the \di\ parameters are
less certain than in other systems.  As in our paper, the authors
assume a constant D/H across the entire absorption complex.  D/H $=
(4.0^{+0.6}_{-0.7}) \times 10^{-5}$.

\subsection{$z=2.536$ towards HS0105+1619}

\citet{OMeara01} analysed this system and found that $b$(\hi) is very
small ($\sim 14$ \kms), meaning that each \di\ line is clearly
separated from the nearby \hi\ absorption.  Consequently $b$(\di) and
$N$(\di) can be measured independently of the \hi\ absorption
parameters. log $N$(\hi) $=19.422 \pm 0.009$ is determined by the damping
wings in the \lya\ line, and \di\ is seen in five transitions. A
relatively precise D/H can be derived: D/H $= (2.54 \pm {0.23}) \times
10^{-5}$.

\subsection{$z=2.076$ towards QSO 2206-199}

\citet{Pettini01a} find a very low $b$(\hi) ($\sim 14$ \kms) and an
extremely simple velocity structure (apparently a single \hi\
component).  This allows a D/H measurement despite the relatively low
S/N of their HST spectra. Two Lyman transitions can be used to
constrain $N$(\di).  log $N$(\hi) $=20.436 \pm 0.008$ is measured
using the damping wings of the \hi\ \lya\ line.  D/H $= (1.65 \pm
0.35) \times 10^{-5}$.

\subsection{$z=3.025$ towards Q0347-3819} 
\label{Lev}
This system was initially analysed by \citet{Dodor01}, who found D/H
$= (2.24 \pm 0.67) \times 10^{-5}$.  It was subsequently re-analysed
by \citet{Levsh02a}, who found D/H $= (3.75 \pm 0.25) \times
10^{-5}$. log $N$(\hi) $=20.626 \pm 0.005$ is measured using the
damping wings in the \lya\ line.  Levshakov et al.'s result is
different because they assume the main component in \hi\ has a similar
$z$ and $b$ to that of H$_2$, also observed in this absorber. The \di\
lines are heavily blended with \hi\ and neither $b$(\hi) nor $b$(\di)
can be measured directly.  The D/H value derived is dependent on the
assumed velocity structure, as shown by the different results of
D'Odorico et al. and Levshakov et al.  The \si4\ and \c4\ transitions
for this absorber in \citet{ProchaskaWolfe99} show components at $\sim
-60$ \kms\ (on the scale in Levshakov et al.'s Fig.~10), which may
have associated \hi\ absorption that could affect the derived D/H.  We
suspect that the range of D/H values allowed in this absorber is
significantly larger than that quoted by \citet{Levsh02a}.  If we were
to include this result, we would choose a conservative $1 \sigma$
error range of $(1.57 - 4.00) \times 10^{-5}$.  Since this range is at
least a factor of $2$ greater than the other D/H error ranges, we
exclude this value from our comparison of D/H values below.

\subsection{$z=2.526$ towards Q1243+3047} 

\citet{Kirkman03} analysed this system.  \di\ is seen in five
transitions, but blended with \hi.  $N$(\hi) is measured using the
damping wings at \lya.  This is complicated by the \lya\ lying close
to the quasar \hi\ \lya\ emission, where the continuum is steep and
difficult to determine.  The authors generate many continua based on
`anchor' points in the spectrum which they assume are free from forest
absorption.  They use these to find a robust estimate for log $N$(\hi)
$=19.73 \pm 0.04$.  The fitted $b$(\di) is marginally consistent with
the predicted value from $b$(\hi) and $b$(\o1).  D/H $=
(2.42^{+0.35}_{-0.25}) \times 10^{-5}$.  
\\ 
\\ 
All of these D/H measurements are in absorbers with metallicities of
[Si/H] $<-1$, and so no significant D astration should have taken
place.  Thus we might expect all of these values,
along with the D/H value from this paper, to be consistent with a
single, primordial D/H.  To test this, we fitted the five adopted D/H
values above together with the value in this paper to a single D/H
value.  Where there are not symmetrical errors around the D/H values,
we force them to be symmetrical by increasing the smaller error to
match the larger error.  We find D/H $= (2.40 \pm 0.3) \times
10^{-5}$, which is consistent with the predicted D/H from CMB and
standard BBN, $(2.60^{+0.19}_{-0.17}) \times 10^{-5}$ \citep{Coc04}.
The value of the total $\chi^2$ for the fit is $25.95$ for five
degrees of freedom.  The chance of exceeding this $\chi^2$ value by
chance is less than 0.01 \%. Thus, taken at face value, the
measurements are not consistent with a single D/H; but with so few
measurements this may just be a result of small number statistics.
\citet{Kirkman03} argue that the inconsistency is probably due to
underestimated systematic errors - indeed, if we artificially double
all the errors of the D/H values, the total $\chi^2 = 6.5$ and the
inconsistency disappears.  However, we note that the statistical
errors on each D/H result were arrived at only after careful
consideration of systematic errors.  This highlights the need for
further D/H measurements to determine whether the scatter in D/H is
real.

\subsection{Correlation between D/H and $N$(\hi)}

\citet{Pettini01a} and \citet{Kirkman03} note that there appears to be
a trend of decreasing D/H with increasing $N$(\hi).  Our result does
not follow this trend, and has a much lower D/H value than either of
the two other low $N$(\hi) systems towards PKS1937-1009 and 1009+2956
(Fig.~\ref{d2hvN}).  We are not sure of the reason for this
difference.  As \citet{Pettini01a} commented, it is interesting that
the two systems with the highest D/H values are both grey LL systems,
where the total $N$(\hi) was measured using the drop in flux at the
LL.  In all the other D/H systems, including the one in this paper,
the dominant constraint on the $N$(\hi) value is from the damping
wings of the \lya\ or \lyb\ lines.  At first glance this might suggest
a systematic error in measuring $N$(\hi) could explain the difference
between our D/H and the grey LL D/H values, although previous authors'
analyses of systematics affecting $N$(\hi) in both grey LL and higher
$N$(\hi) systems have been very thorough.

\section{Discussion}

We have identified a new Lyman limit absorption system towards
PKS1937-1009, with log$N$(\hi) $=18.25 \pm 0.02$ at $z=3.256$.  The
\lya\ and \lyb\ transitions are suitable for measuring D/H, and we
find a $68.3$\% range for D/H of $1.6^{+0.25}_{-0.30}\times 10^{-5}$,
and a $95.4$\% range of $1.6^{+0.5}_{-0.4} \times 10^{-5}$.

The metallicity of the cloud where D/H was measured is low, [Si/H] $=
-2.0 \pm 0.5$.  At this metallicity we expect that the D/H value will
be primordial.

This value disagrees at the $99.4$\% level with the predicted D/H
value using the $\Omega_b$ calculated from the WMAP cosmic background
radiation measurements, $2.60^{+0.19}_{-0.17} \times 10^{-5}$
\citep{Coc04}, and it exacerbates the scatter in D/H measurements
around the mean D/H value. It is consistent with the D/H values
measured in the ISM using FUSE, \citep[e.g.][]{Hebrard03}.  However,
it is not consistent with the idea of a single primordial D/H value
which stays constant until the metallicity reaches $\sim 1/10$ solar,
then begins to drop off, reaching the ISM D/H at solar metallicity.
Some early mechanism for D astration may be the cause for the
scatter in D/H values \citep{Fields01}, although in this case we would
expect to see a correlation between D/H and [Si/H], which is not
observed.

Other authors \citep{ Pettini01a, Kirkman03} note that there is a trend
of decreasing D/H with increasing $N$(\hi) of the absorbers in which
they are measured.  The D/H value in this paper does not follow this
trend, suggesting that the trend is not real and only due to the low
number of D/H measurements.
\\
\\
We would like to thank Antoinette Songaila for making the Keck data
used in this analysis available and Max Pettini for his helpful
correspondence. We also thank the anonymous referee for their
constructive comments that helped improve the paper.  AOG and AFS
acknowledge support from the Generalitat Valenciana ACyT project
GRUPOS03/170, and Spanish MCYT projects AYA2002-03326 and
AYA2003-08739-C02-01.

\bibliographystyle{mn2e}
\bibliography{/home/nhmc/tex/references}

\end{document}